\documentstyle[12pt]{article}
\begin{document}

\pagestyle{plain}

\vspace*{-10mm}

\baselineskip18pt
\begin{flushright}
{\bf BROWN-HET-940(Rev.)}\\
{\bf SNUTP - 94 - 59}\\
{\bf August 1995}\\
\end{flushright}
\vspace{1.0cm}
\begin{center}
\Large
{\bf THE W-BOSON MASS  AND PRECISION TESTS \\ OF THE STANDARD MODEL}\\
\vglue 7mm
\normalsize
{\bf Kyungsik Kang } \\
\vglue 2mm
{\it Department of Physics, Brown University,
Providence, RI 02912, USA,\footnote{Supported
in part by the USDOE contract DE-FG02-91ER40688-Task A.} }\\
\vglue 3mm
{\it and}\\
\vglue 3mm
{\bf Sin Kyu Kang } \\
\vglue 2mm
{\it Department of Physics, Seoul National University, Seoul, Korea
 \footnote{ Supported in part by the Basic Science Research
Institute program, Ministry of Education, 1994, Project No.BSRI-94-2418,
and the Korea Science and Engineering Foundation through
SNU CTP.}}\\
\vglue 10mm
{\bf ABSTRACT} \\
\vglue 8mm
\begin{minipage}{14cm}
{\normalsize
We have examined  the electroweak radiative
corrections in the LEP precision data
in view of the new measurements of $M_W$ and $m_t$ as well as the recent
progress in the higher order radiative corrections.
{}From the minimal $\chi^2$-fit to the experimental Z-decay parameters
(with the aid of a modified ZFITTER program),
we predict that $M_W=80.29(4)(2)$ GeV where the first error is due to
the uncertainty in the fitted $m_t$ for a fixed $m_H$ and the second error
comes from the $m_H$ in the range of $60-1000$ GeV,
which is to be compared with the current world average $M_W=80.23(18)$ GeV.
The current world average value of $M_W$ and the 1994 LEP data
definitely favor nonvanishing electroweak radiative corrections
and are consistent with a heavy $m_t$ as measured by the recent CDF report
but with a heavy Higgs scalar of about 400 GeV within the context  of
the minimal standard model.
The sensitivity of and the errors in the best fit solutions
due to the uncertainties in the gluonic coupling $\alpha_s(M_Z)$ and
$\alpha (M_Z)$ are also studied carefully. In addition
we discuss how the future precision measurements of $M_W$ can provide a
decisive
test for the standard model with radiative corrections
and give a profound implication for the measurement of t-quark and Higgs
masses.
}
\end{minipage}
\end{center}
\newpage
%
%
 Much interests have been paid in recent years to the electroweak radiative
corrections (EWRC)
and precision tests of the standard model thanks to the accurate data
obtained at LEP [1-7].
There have been numerous articles published on the subject as has been
documented in [7-9].
The LEP data are generally   regarded as the support for  the standard model
and as the  evidence of the nonvanishing EWRC [9].

However there have been several important developments since last
year which {\it warrant a new motivation to repeat} the precision tests of the
 standard model.
Some of the experimental advances are (1) the new CDF value for $M_W$
[10], (2) the improved LEP data [11], and (3) the CDF and D0 reports
\footnote {The most recent values [13]
are $m_t = 176 {\pm} 8 {\pm} 10$ GeV (CDF) and
$m_t = 199^{+19}_{-21} {\pm} 22$ GeV (D0). } on $m_t$ [12],
while there have been also some  progress on the higher order corrections,
in particular the dominant two-loop terms of order $\alpha^2 m_t^4$ [14],
the QCD-electroweak mixed diagrams [14,15]
and higher order corrections to the
QCD factor in the $Z$- decay width [16].
We would like to present the results of the new fit to the {\it updated}
1994 data with the aid of the appropriately modified  ZFITTER program
to incorporate these new theoretical developments.
We examine the errors in the best fit solutions due to the uncertainties
in the strong coupling constant $\alpha_s (M_Z)$
and also in $\alpha (M_Z)$.
In the analysis we determine $M_W$ self-consistently
from the W-mass relation that includes EWRC for the value of $m_t$
covering experimental range and fit the LEP data,
and show how stable the predicted $M_W$ is regardless the exact value of
$m_H$ in the interesting range of $60-1000$ GeV.
Though the sensitivity of the EWRC to the exact value of $M_W$
in the standard model has been studied based on the $W$-mass formula [17],
the effect of the self-consistency satisfied by $M_W$ through the mass relation
with EWRC as well as that of the uncertainties in $\alpha_s(M_Z)$ and
$\alpha (M_Z)$ to the precision tests of
the electroweak data has not been fully examined and understood.
For this reason, we would like to critically examine in this paper the
sensitivity of the precision tests and the $m_t-m_H$ correlation to
the requirement of consistency in the needed EWRC for a set of mass values
as well as the errors in $\alpha_s(M_Z)$ and $\alpha (M_Z)$.
 In particular, the results of the minimal $\chi^2$-fit show that the
CDF value $m_t=174$ GeV can be consistent with a best fit solution but with
a Higgs mass about
$m_H $ = 400 GeV if $\alpha_s (M_Z) = 0.123$ and $\alpha^{-1}(M_Z)=128.87$
and predict $M_W=80.29(4)(2)$ GeV where the first error is due to the
uncertainty in the fitted $m_t$ for a fixed $m_H$ and the second error
comes from the $m_H$ in the range $60-1000$ GeV.
However $m_t$ and $m_H$ can easily be shifted, due to the uncertainty
in the gluonic coupling $\Delta \alpha_s = \pm 0.006$,
by as much as 5 GeV and 125 GeV respectively,
while the corresponding shift in $M_W$ turns out to be about $30$ MeV.
The possible shifts in $m_t, m_H$ and $M_W$ due to the error in the gauge
coupling $\Delta \alpha^{-1}(M_Z)=\pm 0.12$ can be as much as 6 GeV, 160 GeV
and 20 MeV respectively.

In addition, we reexamined the claim made by
Novikov, Okun, and Vysotsky [18] that the 1993
data from LEP on the electroweak parameters as defined in the standard model
could be explained by the QED Born approximation (QBA) in which $\alpha (M_Z)$
is used instead of $\alpha(0)$ in the tree approximation
 along with the corresponding redefinition of the weak
mixing angle $\sin ^2\theta $ instead of $\sin ^2\theta_W $.
In particular the so-called QBA predictions were
 claimed to be within $1\sigma $ accuracy of all electroweak precision
measurements made at LEP in 1993.
In order to examine the intriguing claim made in Ref.[18],
we   considered the case  of the QBA by consistently neglecting
the terms of non-photonic origin in the full  EWRC.
The full EWRC is calculated
with the aid of a modified ZFITTER program [19]
that uses an improved QCD correction factor [16],
includes the dominant two-loop and QCD-electroweak
 \footnote{We note that the higher QCD effects ($\alpha \alpha_s^2 m_t^2$
order)
 are yet to be unanimously agreed by the experts as evidenced
by the discussions in [15,20].} terms [14] and
makes the minimal $\chi^2$ fit to the data.
We have found [21] that the situation is very sensitive to the value
of $M_W$ and that the QBA fit to the 1993 data gives statistically comparable
or better $\chi^2$ and therefore the existence of the genuine quantum effect
due to full EWRC might not have been evident in the 1993 data.
We note that the old $M_W$ has somewhat lower central value and larger error.
The new results of the minimal $\chi^2$ fit to the updated 1994 data and
the predicted $M_W$ compared to the new world average $M_W$ show a clear effect
due to the EWRC.

It is well known that the charge renormalization in the conventional QED
fixes the counter term by
the renormalized vacuum polarization $\hat {\Pi}^{\gamma }(0)$ and
one can evaluate
$\hat {\Pi}^{\gamma}(q^2)=\hat {\Sigma} ^{\gamma \gamma }(q^2)/q^2$
from the photon self energy $\hat{\Sigma}^{\gamma \gamma }(q^2)$, for example,
by the dimensional regularization method. This gives at $q^2=M_Z^2$
the total fermionic contribution of $m_f \leq M_Z $ to the real part
$ Re\hat {\Pi}^{\gamma}(M_Z^2) = -0.0596(9)$,
which includes both the lepton and quark parts [22].
Here, the quark contribution to $Re\hat {\Pi }^{\gamma}(q^2)$ is
the hadronic one which can be directly evaluated  by dispersion
integral over the measured cross section of $e^+e^- \rightarrow hadrons$.
Then, we get
$\alpha(M_Z)=1/128.87(12) $ in the on-shell scheme
 if the hyperfine structure constant $\alpha = e^2/4\pi = 1/137.0359895(61)$
is used, which we will use in this paper.
The error in $\alpha(M_Z)$ is due to the uncertainty in hadronic contribution.
This is obviously a source of the uncertainty in the best fit solutions.

The electroweak parameters are evaluated
numerically with the hyperfine structure constant $\alpha$,
the four-fermion coupling constant of $\mu$-decay,
$G_{\mu} = 1.16639(2)\times 10^{-5} \mbox{GeV}^{-2}$, and $Z$-mass
$M_Z = 91.1888(44)$ in the 1994 data fit.
Numerical estimate of the full EWRC requires
the mass values of the leptons, quarks, and Higgs scalar besides these
quantities.
While $Z$-mass is known to an incredible accuracy from the LEP experiments
largely due to the resonant depolarization method, the situation with respect
to the $W$-mass is desired to be improved, i.e., $M_W = 80.22(26)~$ GeV [23]
and 80.23(18) GeV[10] from the world average values  vs.
the old and new CDF measurements
  $M_W = 79.91(39)~$ GeV [24] and $M_W = 80.38(23)$ GeV[10].
The minimal $\chi^2$-fit to the LEP data will at best give the $m_t - m_H$
correlations  as shown in Fig. 1(a) and Fig. 1(b).
We will show how the best fit solutions are chosen out of the solution set
for $(m_t, m_H)$ and determine $M_W$ self-consistently from the $W$-mass
relation that includes EWRC and how they depend on the uncertainties in
$\alpha_s(M_Z)$ and $\alpha (M_Z)$ so that how precisely the standard
model can be tested at the moment, let alone the predictions of the
$m_t - m_H$ correlation from the experimental $M_W$.

One has, in the standard model, the on-shell relation
$\sin^2 \theta_W = 1-\frac{M_W^2}{M_Z^2},$
while the four-fermion coupling constant $G_{\mu}$
can be written as
\begin{equation}
G_{\mu } = \frac{\pi \alpha }{\sqrt{2}M_W^2}
\left(1-\frac{M_W^2}{M_Z^2}\right)^{-1}(1-\Delta r)^{-1}
\end{equation}
so that $\Delta r$, representing the radiative corrections, is given by
\begin{equation}
\Delta r = 1-\left(\frac{37.2802}{M_W}\right)^2
\frac{1}{1-M_W^2/M_Z^2}.
\end{equation}
We note from Table 1 that the radiative correction $\Delta r$
is very sensitive to the value of $M_W$.
Mere change in $M_W$ by $0.59\%$ results as much as a $43\%$ change in
$\Delta r$.
\begin{table}
\begin{center}
\begin{tabular}{|c|c|c|c|}\hline \hline
 & $M_W$ (GeV) & $\Delta r$ & $\sin^2 \theta_W$ \\ \hline
 1 & 79.91 & 0.0623 & 0.2321 \\ \hline
 2 & 80.22 & 0.0448 & 0.2261 \\ \hline
 3 & 80.23 & 0.0443 & 0.2263 \\ \hline
 4 & 80.38 & 0.0355 & 0.2231 \\ \hline
\hline
\end{tabular}
\caption{ Dependence of the radiative correction $\Delta r $
on the values of $M_W$ .}
\end{center}
\end{table}
Theoretically, the radiative correction parameter $\Delta r$ within the
standard
model can be written as [25]
\begin{equation}
1-\Delta r = (1-\Delta \alpha)\cdot(1+\cot^2 \theta_W \Delta \bar{\rho})
            -\Delta r_{rem},
\end{equation}
where $\Delta \bar{\rho }$ contains one loop and the leading 2-loop irreducible
weak and QCD corrections.
Main contribution to $\Delta \bar{\rho }=1-\rho^{-1}$ is from the heavy t-quark
through the mass renormalizations of weak gauge bosons $W $ and $Z,$
while there is a part in $(\Delta r)_{rem}$ containing also the t-quark
and Higgs scalar contributions.
Note that the so-called QBA to $\Delta r$ is defined by keeping only
the photon vacuum polarization contribution,
$\Delta \alpha
= -Re \hat{\Pi }^{\gamma}(M_Z^2)=0.0596$.
We see from Table 1 that $\Delta \alpha $ is numerically the
dominant component of the radiative corrections. In particular for
the old CDF $M_W$, $\Delta \alpha $ is already within $4.4\%$ of the needed
$\Delta r$ and is close enough to be within the
experimental uncertainty.
However, with the current world average value $M_W=80.23 $ GeV,
$\Delta \alpha $ differs by  $32\%$ from the needed $\Delta r$ that has
to be accounted for by the weak interaction corrections [26].

Note that precise determination of the on-shell value of $\sin ^2\theta_W$
can also constrain the needed radiative correction and the value of $M_W$.

We have searched for the minimal $\chi^2$-fits to both 1993 and earlier
1994-data of the Z-decay parameters
measured at LEP by using the modified ZFITTER program [21].
Within the framework of the standard model in which $G_{\mu}, \alpha$ and
$M_Z$ are taken as input, one can predict $M_W$ from (2) in a self-consistent
manner.
Starting with the given masses of the quarks, leptons, gauge bosons and Higgs
scalar, as well as given gluonic coupling $\alpha_s(M_Z)$ and gauge coupling
 $\alpha(M_Z)$,
$\Delta r$ is calculated from (3) by including up to the dominant two-loop and
QCD-electroweak mixed terms and then is used to determine $M_W$ from the
right hand side of (2). With this new $M_W$, $\Delta r$ is calculated again
to determine another new $M_W$.
This iteration process is repeated until $\Delta r $ converges
to within $O(10^{-6})$. The final output $M_W$ from this iteration procedure
is the self-consistent solution of (2) for $M_W$ with the starting set of
$m_t, m_H$ etc.
Upon varying $m_t$, this procedure will give the $m_t-M_W$ correlation
for all other parameters including $m_H$ fixed.
The family of the curves in Fig. 2 represents such correlations for
different fixed values of $m_H$.
The $m_t-m_H$ correlation shown in
Fig. 1(a) and Fig. 1(b) is not restrictive enough to discriminate an
interesting region of $m_t$ and $m_H$ when compared to the experimental
range of $m_t$.
Also the $M_W-m_t$ correlation shown in Fig. 2 does not discriminate $m_H$ as
long as it is heavier than $100$ GeV given the current experimental situation
of $(m_t, M_W)$.
Fig. 3 shows the result of the self-consistent procedure for $M_W-m_H$
correlation when $m_t$ is restricted to the CDF value
$174\pm 16$ GeV on which the predicted $M_W$ range is also indicated.
The error band $\pm 0.04$ GeV in the predicted $M_W$ is due to the uncertainty
in the fitted $m_t$ for a fixed $m_H$ as one can deduce from Table 2.
We then calculate the eleven $Z$-decay parameters, as chosen in
Table 2, for the parameter sets $(m_t, m_H$) that determine $M_W$ from (2)
and search
for the minimal $\chi^2$-fit solution to the experimental $Z$-decay parameters.
This procedure selects the Best.fit curve in Fig. 1 and $\diamond $ points
on each curve in Fig. 2.
Including the error due to varying $m_H$ in the range, we predict
$M_W=80.29(4)(2) $ GeV, which is slightly larger than the central value of
the current world average 80.23 GeV.
Note that the predicted $M_W$ constrains $m_H$ to lie 230 GeV and 830 GeV
for $m_t=174\pm 16 $ GeV as one can see from Fig.3.

The results of the best $\chi^2$-fits to
the updated 1994 data  are given for the four sets of $(m_t, m_H)$ in Table 2.
The $Z$-decay parameters are calculated
with the gluonic coupling constant in the range
$\alpha _s(M_Z) = 0.123(6)$
in the QCD correction factor
$R_{\mbox{QCD}} = 1+1.05\frac{\alpha_s}{\pi}+0.9(\pm 0.1)
\left(\frac{\alpha_s}{\pi}\right)^2-13.0
\left(\frac{\alpha_s}{\pi}\right)^3$
for light quarks(u,d,c,s) [27] and in
$R_{\mbox{QCD}} = 1+c_1(m_b)\frac{\alpha_s}{\pi}+c_2(m_t,m_b)
\left(\frac{\alpha_s}{\pi}\right)^2+c_3(m_t,m_b)
\left(\frac{\alpha_s}{\pi}\right)^3$,
the $m_b$ and $m_t$ mass dependent one for b quarks [16].
The partial width for $Z\rightarrow f\bar{f}$ is given by
\begin{equation}
\Gamma_f = \frac{G_{\mu}}{\sqrt{2}}\frac{M_Z^3}{24\pi}\beta R_{\mbox{QED}}
c_fR_{\mbox{QCD}}(M_Z^2)\left \{ [(\bar{v}^Z_f)^2+(\bar{a}^Z_f)^2]\times
 \left(1+2\frac{m_f^2}{M_Z^2}\right)-6(\bar{a}^Z_f)^2\frac{m_f^2}
{M_Z^2}\right \}
\end{equation}
where $\beta =\sqrt{1-4m_f^2/M_Z^2}$, $R_{\mbox{QED}}
=1+\frac{3}{4}\frac{\alpha}{\pi}Q_f^2$ and the color factor $c_f=3$ for
quarks and 1 for leptons.
Here the renormalized vector and axial-vector couplings are defined by
$\bar{a}_f^Z=\sqrt{\rho_f^Z}2a_f^Z = \sqrt{\rho_f^Z}2I_3^f $ and
$\bar{v}^Z_f=\bar{a}^Z_f[1-4|Q_f|\sin^2\theta_W\kappa^Z_f] $ in
terms of the familiar notations [19,28].
Note that $\Delta \alpha $  is included in the
couplings through $\sin^2 \theta_W$ via (1) and (3) and all other
non-photonic loop corrections are
grouped in $\rho_f^Z$ and $\kappa_f^Z$ as in [19,29]
including the dominant two-loop and QCD-electroweak terms.
Note that the QED loop corrections can unambiguously be separated from
the electroweak loops in the case of neutral current interactions [28].
Thus the case of the QBA
can be achieved simply by setting $\rho^Z_f$ and $\kappa^Z_f$
to 1 in the vector and axial-vector couplings.

\scriptsize
\begin{table}
\begin{center}
\begin{tabular}{|c||c||c||c|c|c|c|} \hline \hline
 & Experiment &Born& Full EW &Full EW& Full EW & Full EW \\
  \hline
 & & & & & & \\
 $m_t$~(GeV) & $174\pm 10^{+13}_{-12}$ & & $187^{(5)(6)}_{(4)(5)}$ &
 $177^{(5)(6)}_{(4)(6)}$
 & $163^{(5)(7)}_{(4)(6)}$ & $149^{(4)(6)}_{(5)(7)}$ \\
 & & & & & & \\
 $ m_H$~(GeV) & 60 $\leq m_H \leq 1000$ && $1000$ &
 $500$ & 200 & 60 \\
 & & & & & &\\
 \hline
 & & & & & &\\
$M_W$~(GeV) & $80.23\pm 0.18$ &$79.96^{(2)}_{(1)}$
& 80.31(3)(2) &$80.30^{(2)}_{(3)}(2)$ &
$80.28(3)^{(3)}_{(2)}$ & $80.27(3)^{(2)}_{(3)}$ \\
 & & & & & & \\
 $ \Gamma_Z $~(MeV) & $2497.4\pm 3.8 $ &$2489.0^{(0.9)}_{(0.8)}$
& $2496.5^{(2.0)(0.8)}_{(1.8)(0.6)}$ &
$2496.7^{(1.9)(0.8)}_{(2.1)(0.8)}$ &
$ 2496.5^{(2.2)(1.0)}_{(1.9)(0.7)}$ &
$2495.7^{(2.4)(0.6)}_{(2.2)(0.8)}$ \\
 & & & & & & \\
$ \sigma_{h}^P(nb)$ & $41.49\pm 0.12 $&$41.41^{(0)}_{(0)}$
& $41.40^{(3)(1)}_{(3)(0)}$ &
$41.39^{(3)(1)}_{(3)(0)}$& $41.39^{(3)(0)}_{(3)(1)}$ &
$ 41.38^{(3)(0)}_{(3)(0)}$ \\
& & & & & & \\
$R(\Gamma_{had}/\Gamma_{l\bar{l}})$ & $20.795 \pm 0.040$ &
$20.850^{(5)}_{(6)}$& $20.789^{(39)(7)}_{(39)(6)}$
 & $20.798^{(40)(7)}_{(38)(6)}$ & $20.809^{(38)(7)}_{(39)(6)}$ &
$ 20.822^{(39)(6)}_{(39)(7)}$ \\
 & & & & & & \\
$ A^{0,l}_{FB}$ & $ 0.0170\pm 0.0016 $&$0.0168^{(5)}_{(6)}$&
$ 0.0148^{(4)(3)}_{(2)(1)}$ &
$0.0149^{(2)(2)}_{(3)(2)}$& $0.0149^{(3)(2)}_{(2)(1)}$ &
$0.0152^{(2)(2)}_{(2)(2)}$ \\
& & & & & & \\
$ A_{\tau} $ & $ 0.143\pm 0.010$&$0.150^{(2)}_{(3)}$&
$0.141^{(1)(1)}_{(1)(1)}$ &
$0.141^{(1)(1)}_{(1)(1)}$&
$0.141^{(1)(1)}_{(1)(1)}$ & $0.142^{(1)(1)}_{(1)(1)}$ \\
& & & & & & \\
$ A_{e} $ & $ 0.135\pm 0.011$&$0.150^{(2)}_{(3)}$&
$0.141^{(1)(1)}_{(1)(1)}$ &
$0.141^{(1)(1)}_{(1)(1)}$&
$0.141^{(1)(1)}_{(1)(1)}$ & $0.142^{(1)(1)}_{(1)(1)}$ \\
& & & & & & \\
$R(\Gamma_{b\bar{b}}/\Gamma_{had})$ & $0.2202 \pm 0.0020$ &
$0.2180^{(0)}_{(0)}$&$0.2154^{(1)(2)}_{(0)(1)}$
&$ 0.2158^{(1)(1)}_{(1)(2)}$&$ 0.2161^{(1)(2)}_{(0)(1)}$ &
 $0.2165^{(1)(2)}_{(0)(2)}$ \\
 & & & & & &\\
$R(\Gamma_{c\bar{c}}/\Gamma_{had})$ & $0.1583 \pm 0.0098$ &
$0.1707^{(0)}_{(0)}$&$0.1711^{(0)(0)}_{(0)(0)}$
& $0.1710^{(1)(1)}_{(0)(0)}$&
$ 0.1709^{(1)(1)}_{(0)(0)}$ &
$ 0.1709^{(0)(0)}_{(0)(1)}$ \\
& & & & & & \\
$ A^{0.b}_{FB}$ & $  0.0967\pm 0.0038$&$0.1050^{(17)}_{(18)}$&
$0.0985^{(11)( 8)}_{(8)(5)}$ &
$0.0987^{(10)( 6)}_{(8)(6)}$ & $0.0989^{(10)(7)}_{(8)(4)}$ &
$0.0998^{(8)(6)}_{(7)(7)}$ \\
& & & & & & \\
$ A^{0.c}_{FB}$ & $  0.0760\pm 0.0091$&$0.0750^{(14)}_{(14)}$&
$0.0701^{(9)(6)}_{(6)(4)}$ &
$0.0703^{(6)(4)}_{(8)(5)}$& $0.0704^{(7)(5)}_{(6)(3)}$ &
$0.0711^{(6)(4)}_{(5)(5)}$ \\
& & & & & & \\
$ \sin^2\theta^{lepton}_{eff}$ & $0.2320\pm 0.0016 $
&$0.2321^{(3)}_{(3)}$ & $0.2327^{(1)(1)}_{(2)(2)}$ &
$0.2326^{(1)(1)}_{(2)(2)}$ &
 $ 0.2324^{(1)(1)}_{(2)(1)}$ &
 $0.2322^{(1)(1)}_{(2)(2)}$ \\
& & & & & & \\
\hline
$\chi^2 $ &&19.5 & 11.2 &10.3& 9.57 & 9.31 \\
\hline
 & & & & && \\
$ \Delta r$ & $  0.0443\pm 0.0102$ &$0.0596^{(9)}_{(9)}$
&$ 0.0397^{(16)(10)}_{(19)(13)}$  &
$0.0403^{(19)(12)}_{(16)(13)}$& $0.0414^{(14)(11)}_{(19)(16 )}$ &
$0.0422^{(14)(14)}_{(14)(12)}$ \\
& & & & & & \\
\hline \hline
\end{tabular}
\normalsize
\caption{Numerical results including full EWRC for
eleven experimental parameters of the Z-decay and $M_W$.
Each pair of $m_t$ and $m_H$ represents the case of the best $\chi ^2$-
fit to the 1994 LEP data for $\alpha_s(M_Z)
=0.123(6) $ and $\alpha^{-1}(M_Z)=128.87(12)$.
The numbers in () represent the errors due to $\Delta \alpha_s(M_Z) =
\pm 0.006$ and $\Delta \alpha^{-1}(M_Z)=\pm 0.12$ respectively.
$\sin^2\theta^{lepton}_{eff}$ is from the measurement of $<Q_{FB}>$.
For the case of Born approxiamtion, the errors are
due to $\Delta \alpha^{-1}(M_Z)$ only.}
\end{center}
\end{table}
%
\normalsize
\baselineskip18pt
Numerical results for the best $\chi^2$-fits to the 1993 LEP experimental
parameters of $Z$-decay
for $M_W = 79.91(39) $ GeV [24] and $ M_W = 80.22(26)$ GeV [23]
showed [21] generally small
contributions of the weak corrections
and in particular that the QBA was close to the experimental values
within the uncertainty of the measurements, i.e., within $2\sigma$.
The near absence of the weak interaction contributions to the
radiative corrections for the 1993 data is more impressive for
$M_W$ = 79.91(39) GeV
 than for $M_W$ = 80.22(26) GeV. This is mainly because QBA gives
 $M_W$ = 79.95 GeV
compared to $M_W$ = 80.10(1) GeV from the full EWRC for $m_H$ in the range
of $60 - 1000$ GeV and a larger error in the former as observed in [18,21].
At closer examination, however, the QBA
in this case over-estimates the radiative corrections and the full
EWRC fair better; for $M_W$ = 80.22(26) GeV and $m_H$ = $60 - 1000$ GeV,
$\Delta r$ = 0.0596 in QBA and $0.0498 - 0.0505$
in the full EWRC to be compared to the required value 0.0448. Also
the global fit to the 1993 data with two variables $m_t$ and $m_H$
in the range $60-1000 $ GeV show
[21] that the best fits can be  achieved by $m_t = 139(17) $ GeV
wth a stable output $M_W=80.12(1)~$ GeV when the full EWRC are
taken into account.

The situation with the minimal $\chi^2$-fits to the updated 1994 LEP data
and with theoretically determined $M_W$ is significantly
different from the case of the 1993 data as one can see from Table 2.
Not only there is clear evidence of the full EWRC in each of the eleven
$Z$-parameters
but also the best fit solutions to the 1994 data show
a stable output $M_W = 80.29(2) $ GeV for $m_H$ in the range of
$60 - 1000$ GeV.
In particular
the QBA gives distinctively inferior $\chi^2$(=19.5/11) for the  1994 data.
Also the CDF $m_t=174$ GeV is a possible output solution
with a $m_H $ about 400  GeV
among the many possible combinations of $(m_t, m_H)$ given by the 'Best.fit'
 curve in Fig. 1(a) and Fig. 1(b).
As shown in Table 2, the Best.fit solutions can have errors due to the
uncertainty in $\alpha_s(M_Z)$ : $m_t$ and $M_W$ may be shifted by as much as
$\pm 5$ GeV and $\pm 30 $ MeV respectively because of
$\Delta \alpha_s=\pm 0.006$.
The error range of the Best.fit solutions is indicated by the curves A and B
in Fig. 1(a).
There are additional comparable errors due to the uncertainty in
$\alpha (M_Z)$ as shown in Table 2 and Fig. 1(b) : $\Delta \alpha^{-1}(M_Z)=\pm
0.12$ can cause another $\pm 6$ GeV
and $\pm 20 $ MeV respectively in $m_t$ and $M_W$.
In general the $\chi^2$-value tends to prefer the lower $m_t$
 and accordingly smaller $m_H$,
though  any pair of $(m_t, m_H)$ on the Best.fit curve in Fig. 1(a)
and Fig. 1(b) is
 statistically comparable to each other.
In particular the best global fits to the updated 1994 data give
$m_t = 155 - 187$ GeV for $m_H = 100 - 1000$ GeV.
Most of the $Z$-parameters are stable irrespectively to the uncertainties due
to $\Delta \alpha_s $ and $\Delta \alpha $ and in excellent agreement with
the data except
$R(\Gamma_{b\bar{b}}/\Gamma_{had})$. Even with
the mass dependent QCD factor, there is still about 2.4 $\sigma$ deviation in
$R(\Gamma_{b{\bar b}}/\Gamma_{had})$ from the experiments irrespectively to
the uncertainties in $\alpha_s(M_Z)$ [15].
 Most of the $\chi^2 $ contributions are from $R(\Gamma_{b\bar{b}}/
\Gamma_{had})$ and to a lesser degree from $R(\Gamma_{c\bar{c}}/\Gamma_{had})$
and $A^{0,l}_{FB}$.
Fig. 2 shows how $M_W$ changes with $m_t$ for fixed $m_H$ from  the consistency
of the full EWRC, on which the new world average $M_W$ and CDF
 $m_t$ along with the Best.fit solutions (depicted by $\diamond $ points)
are also shown.
The central values of the world average  $M_W$ and CDF $m_t$ are
 consistent with a Higgs scalar mass about 1000 GeV,
though $m_H = 100$ GeV is within 1 $\sigma$ because of large errors in the
data.
Clearly a better precision measurement of $M_W$ is desired to distinguish
 different $m_H$.
For example, a change of $m_H$ by 200 GeV, i.e., from 400 GeV to 200 GeV
at $m_t=174$ GeV, results a change of 50 MeV in $M_W$, i.e.,
from 80.30 GeV to 80.35 GeV, as one can see from Fig. 2 and Fig. 3 .
This in turn will require a
 precision of 11 GeV or better in $m_t$ from the Best.fit curve in Fig. 1(a),
which is consistent with the most statistical error improvement that may be
achieved at the Fermilab Tevatron.
Present precisions in the data entail a theoretical uncertainty of about
36 MeV in $M_W$ which is about the overall error improvement expected at
LEP-200.

We have examined the results of the minimal $\chi^2$-fits to the precision
measurements of the Z-decay parameters at LEP
with the aid of a modified ZFITTER program containing the full
one-loop and dominant two-loop EWRC.
While the result of QBA might appear to be
 in agreement with the 1993 data within $2\sigma$ level of
accuracy [21,30], the new world
average value of $M_W$ and updated 1994 LEP data definitely disfavor the QBA
and support for the evidence of the nonvanishing weak correction.
Furthermore, the CDF $m_t$ is a solution of the minimal $\chi^2$-fits to
the 1994 data
with a Higgs scalar mass {\it about} 400 GeV. However this $m_t$ value can be
shifted by as much as 8 GeV due to the overall uncertainties in $\alpha_s(M_Z)$
and $\alpha (M_Z)$ for the moment and accordingly $m_H$ ranging $250 - 690$
GeV.
Further precision measurement of $M_W$ can provide
a real test of the standard model as it will give a tight constraint for the
needed amount of the EWRC and can provide a profound implication for the mass
of t-quark and Higgs scalar.
 If $M_W$ is determined to within a 40 MeV uncertainty,  $\Delta r$ within the
context of the standard model will be tightly constrained to distinguish
 the radiative corrections and the $\chi^2$-fit to the Z-decay data with the
 1994 accuracy
 can discriminate the mass range of the t-quark and Higgs scalar within 8 GeV
 and 200 GeV respectively, providing a crucial test for and even the need of
 new physics beyond the standard model.
Finally if $M_W$ is determined to be larger than $80.31$ GeV with
better than a $20$ MeV accuracy by the future precision measurements
(perhaps reachable at LHC),
 this would be a definite sign for new physics beyond the standard model.

{\it Note added}: Most recent CDF
measurement is $M_W = 80.41(18)$ GeV [31].
We note that the theoretical prediction of $M_W$, 80.29(4)(2) GeV,
is about 0.67$\sigma $ below the central value of the new experimental $M_W$.
\section*{Acknowledgements}
One of us (KK) would like to thank the Center for
Theoretical Physics, Seoul National University (CTPSNU),
the Korea Advanced Institute of Science and Technology (KAIST)
 and LPTPE, Universit\'e P.\&M. Curie
 where parts of the work were done,
for the kind hospitality during his sabbatical
stay. Also the authors would like to thank
Professors Hi-sung Song, Jae Kwan Kim, R. Vinh Mau and other colleagues
at CTPSNU, KAIST, and LPTPE for the stimulating environment and supports
and in particular Professor M. Lacombe
for checking the numerical computations
and to Professor J.E.Kim for his encouragements.

\section*{References}
\begin{description}
\item[1.] ALEPH Collab., D. Buskulic et al., CERN-PPE/93-40 (1993).
\item[2.] DELPHI Collab., D. Aarnio te al., Nucl. Phys. {\bf B367}
          (1911) 511.
\item[3.] L3 Collab., O. Adriani et al., CERN-PPE/93-31 (1993).
\item[4.] OPAL Collab., P. D. Action et al., CERN-PPE/93-03 (1993).
\item[5.] The LEP Collab., CERN-PPE/93-157 (1993).
\item[6.] C. DeClercq; V. Innocente; and
          R. Tenechinl, in: Proc. XXVIIIth Rencontres de Moriond
          (Les Arcs, 1993).
\item[7.] L. Rolandi, in: Proc. XXVI ICHEP 1992, CERN-PPE/92
          -175 (1992); M. P. Altarelli, talk given at Les Rencontres
          de Physique de la Vallee d'Aoste (La Thuil, 1993),
          LNF-93/019(p); and J. Lefranceis, in: Proc.
          Int. EPS. Conf. H. E. Phys. (Marseille, July 1993).
\item[8.] F. Dydak, in: Proc. 25 Int. Conf. H. E. Phys.,
          Eds. K. Phua, Y. Tamaguchi (World Scientific, Singapore, 1991), p3.
\item[9.] W. J. Marciano, Phys. Rev. {\bf D 20} (1979) 274;
          A. Sirlin, Phys. Rev. {\bf D22} (1980) 971; {\bf D29} (1984) 89;
          A. Sirlin and W. J. Marciano, Nucl. Phys. {\bf B189} (1981) 442;
          and A. Sirlin, NYU-TH-93/11/01. See also W. Hollik, in:
          {\it Precision Tests of the Standard Model}, ed. P.
          Langacker (World Scientific Pub., 1993); G. Altarelli, in: Proc.
          Int.EPS. Conf. H. E. Phys. (Marseille, July 1993);
           K. Hagiwara, S. Matsumoto, D. Haidt and C. S. Kim,
	   KEK-TH-375, to be published in Z. Phys. C; and
           J. Erler and P. Langacker, UPR-0632T.
\item[10.] R. M. Keup, Fermilab-Conf-94/282-E; C. K. Jung, in: Proc. 27th
           ICHEP (Glasgow, July 1994).
\item[11.] P. Clarke; Y. K. Kim; B. Pictrzyk; P. Siegris; and M. Woods, in:
           Proc. 29th Rencontres de Moriond (Meribel,1994);
           R. Miquel, in : Proc. 22nd INS Symposium (Tokyo, March 1994); and
           D. Schaile, in: Proc. 27th ICHEP (Glasgow, 1994).
\item[12.] F. Abe et al., Phys. Rev. Lett. {\bf 73} (1994) 225; Phys. Rev.
           {\bf D50} (1994) 2966; S. Abachi et al, Phys. Rev. Lett. {\bf 72}
           (1994) 2138; FERMILAB-PUB-94/354-E.
\item[13.] F. Abe et al., FERMILAB-PUB-95/022-E; S. Abachi et al., FERMILAB-
           PUB-95/028-E.
\item[14.] B. A. Kniehl, Int.J.Mod.Phys.{\bf A10} (1995) 443.
\item[15.] B. H. Smith and M. B. Voloshin, UMN-TH-1241/94; S. Franchiotti, B.
           Kniehl and A. Sirlin, Phys. Rev. {\bf D 48} (1993) 307;
           M. Shifman, TPI-MINN-94/42-T.
\item[16.] K. G. Chetyrkin, J. H. Kuhn, and A. Kwiatkowski, TTP94-32 and
           hep-ph/9503396.
\item[17.] Z. Hioki, Z. Phys. {\bf C49} (1991) 287; Phys. Rev. {\bf D45} (1992)
	   , 1814 ;Mod. Phys. Lett. {\bf A7} (1992) 1009; K. Kang, in: Proc.
           14th Int. Workshop Weak Interactions and Neutrinos (Seoul,
           1993), Brown-HET-931 (1993); and
           K. Kang and S. K. Kang, in: Proc. Workshop on Quantum Infrared
           Physics (Paris, June 1994), Brown-HET-968 and SNUTP-94-97.
\item[18.] V. A. Novikov, L. B. Okun and M. I. Vysotsky, Mod. Phys. Lett.
          {\bf A8} (1993) 5929. See, however,  CERN-TH-7214/94 for more recent
          analysis.
\item[19.] D. Bardin et al., CERN-TH-6443-92 (1992).
\item[20.] T. Takeuchi, A. K. Grant and M. P. Worah, FERMILAB-PUB-94/303-T.
\item[21.] K. Kang and S. K. Kang, in : Proc. Beyond the Standard Model IV
          (Granlibakken, Lake Tahoe, 1994), Brown HET-979 and  SNUTP-94-128
          (December, 1994).
\item[22.] F. Jegerlehner, PSI-PR-91-08 (April 1991).
\item[23.] Particle Data Group, Review of Particle Properties,
          Phys. Rev.{\bf D 45}, No.11, Part II (1992).
\item[24.] CDF Collab., F. Abe et al., Phys. Rev. {\bf D 43} (1991) 2070.
\item[25.] W. Hollik (Ref. 7), p49.
\item[26.] K. Kang (Ref. 17); and Z. Hioki, Mod. Phys. Lett.
	  {\bf A 10} (1995) 121.
\item[27.] T. Hebbeker, Aachen preprint PITHA 91-08 (1991); and
           S. G. Gorishny, A. L. Kataev and S. A. Larin, Phys. Lett. {\bf B
259}
           (1991) 144; and L. R. Surguladze and M. A. Samuel,
           Phys. Rev. Lett. {\bf 66} (1991) 560.
\item[28.] See, for example, W. Hollik (Ref. 9); CERN Yellow Book CERN 89-08,
	   vol.1, p45 ; and K. Kang(Ref. 17).
\item[29.] M. Consoli and W. Hollik, in {\it Z Physics at LEP 1}, Vol. 1,
           eds. G. Altarelli et al., CERN 89-08 (1989).
\item[30.] See also K. Hagiwara, et.al. (Ref.9).
\item[31.] F. Abe {\it et al.}, Phys. Rev. Lett. {\bf 75} (1995) 11.
\section*{Figure Captions}
\item[Fig. 1(a)]: The mass ranges of $m_t$ and $m_H$ from the minimal
$\chi^2$-fit
 to the updated 1994 LEP data for
$\alpha_s(M_Z)=0.123(6)$ and $\alpha^{-1}(M_Z)=128.87(12)$.
The range of the Best.fit solutions (see the text) due to the error
$\Delta \alpha_s=\pm 0.006$ is also indicated by the curves A and B.
\item[Fig. 1(b)]: The mass ranges of $m_t$ and $m_H$ from the minimal
$\chi^2$-fit
 to the updated 1994 LEP data for
$\alpha_s(M_Z)=0.123(6)$ and $\alpha^{-1}(M_Z)=128.87(12)$.
The range of the Best.fit solutions due to the error
$\Delta \alpha^{-1}(M_Z)=\pm 0.12$ is also indicated by the curves a and b.
\item[Fig. 2]: $M_W$ versus $m_t$ for fixed values of $m_H$ from the full
 radiative correction in the standard model.
 The case of the minimal $\chi^2$-fit to the updated 1994 LEP data
 are indicated by $\Diamond $.
\item[Fig. 3]: $M_W$ versus $m_H$ for $m_t=174\pm 16$ GeV :
The middle region bounded by the solid lines represent the predicted
theoretical value of $M_W$.
 Notice that as $m_t$ becomes heavier than 174 GeV the lower bound of
 $m_H$ increases above 300 GeV from the predicted $M_W$.
\end{description}
\end{document}